\documentclass{appolb}
\usepackage{amsmath}
\usepackage{graphicx}
\usepackage{psfrag}

\newcommand{\dl}{\Delta l}
\newcommand{\de}{\Delta\ev}
\newcommand{\zu}{z_u}
\newcommand{\zd}{z_d}
\newcommand{\ev}{\mathbf{e}}
\newcommand{\tev}{\widetilde{\textbf{e}}}
\renewcommand{\e}{\mathbf{e}}
\newcommand{\R}{\bar{R}}
\newcommand{\te}{\widetilde{e}}
\newcommand{\tzu}{\widetilde{z_u}}
\newcommand{\tzd}{\widetilde{z_d}}
\newcommand{\tdl}{\widetilde{\Delta l}}
\newcommand{\tz}{\widetilde{z}}
\newcommand{\ty}{\widetilde{y}}
\newcommand{\tphi}{\widetilde{\theta}}
\newcommand{\ttheta}{\widetilde{\theta}}
\newcommand{\tT}{\widetilde{T}}

\newcommand{\dt}{\Delta t}
\newcommand{\dth}{\Delta \theta}
\newcommand{\dy}{\Delta y}
\newcommand{\dz}{\Delta z}
\newcommand{\avg}[1]{\left\langle #1\right\rangle}
\renewcommand{\d}{\text{d}}

\begin{document}
\title{System response kernel calculation for list--mode reconstruction in strip PET detector
  \thanks{Presented at Symposium on Applied Nuclear Physics and Innovative Technologies, 3-6 June 2013, Krak\'{o}w, POLAND}%
} 
\author{P.~Bia\l{}as$^{1}$, J.~Kowal$^{1}$, A.~Strzelecki$^{1}$, T.~Bednarski$^{1}$, E.~Czerwi\'{n}ski$^{1}$,
\L{}.~Kap\l{}on$^{1}$, A.~Kochanowski$^{1}$, G.~Korcyl$^{1}$, P.~Kowalski$^{2}$, T.~Kozik$^{1}$, W.~Krzemie\'{n}$^{1}$, 
M.~Molenda$^{1}$, P.~Moskal$^{1}$, Sz.~Nied\'{z}wiecki$^{1}$, M.~Pa\l{}ka$^{1}$, M.~Pawlik$^{1}$, 
L.~Raczy\'{n}ski$^{2}$, 
Z.~Rudy$^{1}$, P.~Salabura$^{1}$, N.G.~Sharma$^{1}$, M.~Silarski$^{1}$, A.~S\l{}omski$^{1}$, J.~Smyrski$^{1}$, 
W.~Wi\'{s}licki$^{2}$, M.~Zieli\'{n}ski$^{1}$
\address{$^{1}$~Jagiellonian University, 30-059 Krak\'ow, POLAND\\
         $^{2}$~{\'S}wierk Computing Centre, National Centre for Nuclear Research, 05-400~Otwock-{\'S}wierk, POLAND}
\address{}
}

\maketitle{}
\begin{abstract}
  Reconstruction of the image in Positron Emission Tomographs (PET)
  requires the knowledge of the system response kernel which describes
  the contribution of each pixel (voxel) to each tube of response
  (TOR). This is especially important in list--mode reconstruction
  systems where an efficient analytical approximation of such function
  is required.  In this contribution we present a derivation of the
  system response kernel for a  novel 2D strip PET.
\end{abstract}

\section{Introduction}

The Positron Emission Tomograph (PET) works by estimating the
radioactive fluid density (tracer) from the measurements of the
$\gamma$ quanta emitted from the beta plus ($\beta^{+}$) decay. The two quanta are
emitted simultaneously and almost back-to-back. We will call such
emission an {\em event}. The $\gamma$ are detected in the detectors
surrounding the patient. Detecting two quanta yields a {\em tube of
  response} passing through the emission point. The better the spatial
resolution of the detection the thinner is the tube giving a better
reconstruction. Currently all PET scanners perform the measurements
using the non-organic scintillating crystals and the spatial resolution is
controlled by the crystal size which can be as small as few
millimeters across.

Our group is working currently on a prototype PET using the long
plastic scintillator strips where the spatial resolution is obtained
from the time of flight
measurements~\cite{PMbio11,stripPET,organicscint}. Achieving sufficient time
resolution (less than $100$~ps) is the main technological challenge,
however the novel hardware require also the suitable adaptation of the
reconstruction algorithm.

This contribution is concerned with the calculation of the system
kernel in the 2D image reconstruction in the axial plane of our strip
PET detector. It is organized as follows: in
section~\ref{sec:detector} we describe the detector geometry and
measurment errors, in section~\ref{sec:LM} the principles of the
List-Mode Expectation Maximization Algorithm is described and in 
the following sections we derive the system response kernel.

\section{Detector geometry}
\label{sec:detector}

\begin{figure}
\begin{center}
\psfrag{yl}{$y$}
\psfrag{zl}{$z$}
\psfrag{th}{$\theta$}
\psfrag{xy}{$(y,z)$}
\psfrag{zu}{$\zu$}
\psfrag{zd}{$\zd$}
\includegraphics[width=10cm]{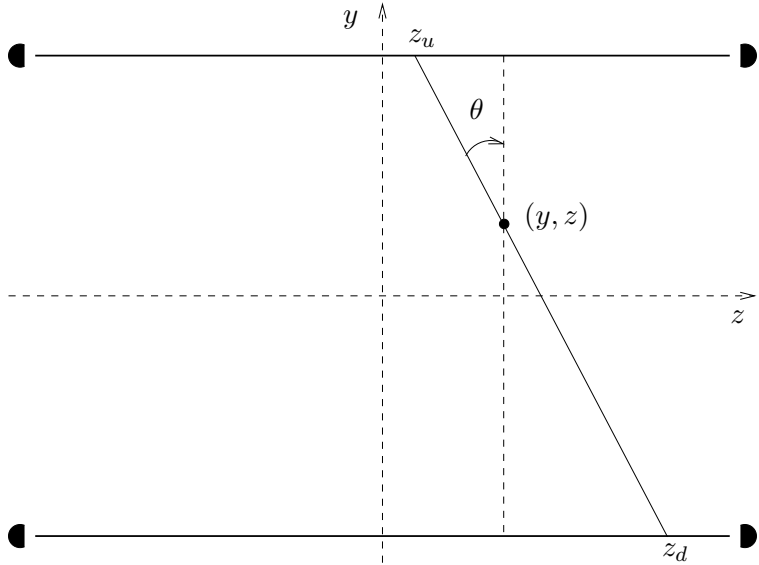}
\end{center}
\caption{\label{fig:detector} Detector geometry.}
\end{figure}

In its final form our detector should consist of strips of
scintillators arranged on a cylinder. The strips are aligned with the
axis of the cylinder. 
 We will start with a simpler 2D geometry --
two parallel line segments of scintillators of length $L$  at the distance $2\R$ (see
figure~\ref{fig:detector}). 
This is anyway a necessary step as our first prototype will consist of
two bars of scintillators. This is in a sense a minimal configuration
required for testing. The real idealization here is neglecting the
scintillator thickness.

A photomultiplier tube is attached to the end of each strip. The
$\gamma$ quanta can scatter in the scintillator and produce light
which then propagates along the scintillator to the
photomultipliers. By measuring the time at which light reaches the
photomultiplier we can estimate the position at which $\gamma$ had
crossed the scintillator 
\begin{equation}
\tzu = \frac{1}{2}c_{sci.}\left(\tT_{ul}-\tT_{ur}\right),			
\quad
\tzd = \frac{1}{2}c_{sci.}\left(\tT_{dl}-\tT_{dr}\right)
\end{equation}
We use tildas to mark the measured
quantities as opposed to the exact  ones.
The $c_{sci.}$ denotes the effective speed of light in the
scintillator. It takes into account both the actual speed of light in
scintillator and the elongation of the optical path due to
reflections. We have estimated this to be approximately $1.3\times
10^{8}$m/s for the scintillators we use.  Combining the time
measurements from the two scintillators we can estimate the position
of the emission point on the line joining the upper and lower crossing
points
\begin{equation}
\tdl=\frac{1}{2}c\left(\left(\tT_{ul}+\tT_{ur}\right)-\left(\tT_{dl}+\tT_{dr}\right)\right)
\end{equation}
where $\dl$ is the difference of distances of the reconstructed point
$(y,z)$ from the upper and lower detection points (see
figure~\ref{fig:detector}).

Those quantities are of course subject to measurement errors and are
related to exact ones by
\begin{equation}
  \tz_{y}=z_{y}+\varepsilon_{z_y},\,y=u,d\qquad \tdl=\dl+\varepsilon_{\dl}.
\end{equation}

We assume that the errors $\varepsilon$ are normally distributed with
some correlation matrix $C$. In general the magnitude of the errors
will depend on place where the $\gamma$ hit the scintillator
$C=C(\zu,\zd)$. This matrix is a necessary and important input for the
reconstruction algorithm. Under some plausible assumptions which are
beyond the scope of this contribution this matrix can be parametrized
by three functions
\begin{equation}
C=\begin{pmatrix}
\sigma_z^2(z_u) & 0    & \gamma(z_u)  \\
0    & \sigma_z^2(z_d) & -\gamma(z_d)\\
\gamma(z_u) & -\gamma(z_d)       &\sigma_{\dl}^2(z_u,z_d)
\end{pmatrix}
\end{equation}
where
\begin{equation}
\sigma^2_z(z)=\avg{\varepsilon^2_{u(d)}(z)},\qquad
\sigma_{\dl}^2(\zu,\zd)=\avg{\varepsilon_{\dl}^2(\zu,\zd)}
\end{equation}
and 
\begin{equation}
  \gamma(z)=\avg{\varepsilon_{\zu}(z)\varepsilon_{\dl}(z,\zd)}
=-\avg{\varepsilon_{\zd}(z)\varepsilon_{\dl}(\zu,z)}
\end{equation}

The $\zu$, $\zd$ and $\dl$ are related to the coordinates $(y,z)$ of
the emission point and the emission angle $\theta$ by the formulas
\begin{equation}
\begin{aligned}
\zu& =   z+(\R-y)\tan\theta\\
\zd& =   z-(\R+y)\tan\theta\\
\dl& = -2 y \sqrt{1+\tan^2\theta}
\end{aligned}
\end{equation}
and  conversely
\begin{equation} 
\begin{aligned}
\tan \theta &= \frac{\zu-\zd}{2\R}\\
y &=-\frac{1}{2}\frac{\dl}{\sqrt{1+\tan^2\theta}}
=\frac{2\R\dl}{\sqrt{\zu-\zd+4\R^2}}\\
z&=\frac{1}{2}\left(\zu+\zd+2y\tan\theta\right)
= \frac{1}{2}\left(
\zu +\zd +\frac{(\zu-\zd)\dl}{\sqrt{\zu-\zd+4\R^2}}
\right).
\end{aligned}
\end{equation}

\section{List--Mode reconstruction}
\label{sec:LM}

Given good enough time resolution our detector using the
time-of-flight technique could reconstruct each individual event with
sufficient accuracy to measure the emitter density directly.
Currently however this is not the case and the measurements errors
have to be incorporated into the reconstruction  using a {\em
  statistical} approach. Almost every current reconstruction algorithm
is based on likelihood maximization approach described
in\cite{sheppvardi,sheppvardikaufman}. This work is concerned with
{\em binned} data. However because of the advance of the technology
most of the scanners work in the list--mode where every single
detected event is recorded separately. The extensions of the
likelihood maximization approach to this case was done
in\cite{lmlh,lmalg}.

Here we provide a very brief introduction to this algorithm, for
details reader is referred to \cite{lmalg}.  Let's denote the {\em
  system response kernel} by $P(\tev|i)$. This is defined as
probability that a {\em detected} event emitted from pixel $i$ was
reconstructed as $\tev$. Given this probability for each emitter
density $\rho$ we can calculate the probability of observing the
particular set of $N$ events\cite{lmlh}:
\begin{equation}
P(\{\tev_1,\dots,\tev_N\}|\rho)=\prod_j\sum_{i}P(\tev_j|i)\frac{\rho(i)s(i)}{\sum_i\rho(i)s(i)}
\end{equation} 
The $s(i)$ is the {\em sensitivity} of the pixel \eg the probability
that an event originating from pixel $i$ will be detected at
all. Together $s(i)$ and $P(\{\tev_j\}|\rho)$ provide the complete
model of the detector.

The reconstruction algorithm consists of finding the distribution
$\rho$ that maximizes this probability, or more accurately its
logarithm - the likelihood.   That is achieved using the iterative
Expectation Maximization (EM) algorithm \cite{lmalg}
\begin{equation}\label{eq:iter}
\rho(l)^{(t+1)}=\sum_{j=1}^N
\frac{P(\tev_j|l)\rho(l)^t}
{T\sum\limits_{i=1}^M P(\tev_j|i)s(i)\rho(i)^t}.
\end{equation}
The sum over $j$ runs over all collected events
$\{\te_j\}$. Considering that up to hundred millions of events
can be collected during a single scan this is a very time consuming
calculation. Finding an efficient approximation for the system
response kernel is of a paramount importance.

\section{System response kernel}

To calculate $P(\te|i)$ we start with $p(\tev|\ev)$ -- the probability that a
event $\ev$ will be detected as $\tev$. This includes the possibility
of an event not being detected
\begin{equation}
 s(\ev)\equiv\int\d\tev\;p(\tev|\ev) \le 1.
\end{equation}
The $s(\e)$ is the sensitivity of an event -- the probability that the
event will be detected. 

With this  definition
\begin{equation}\label{eq:P}
P(\tev|i)=\frac{p(\tev|i)}{s(i)}
\end{equation}
where 
\begin{equation} 
p(\tev|i)=\pi^{-1}\int\limits_{y,z\in i}\int\text{d}\theta p(\tev|(y,z,\theta))
\end{equation}
and
\begin{equation}\label{eq:s}
s(i)=\pi^{-1}\int\text{d}\tev p(\tev|i)=\int\limits_{y,z\in i}\int\text{d}\theta s(y,z,\theta).
\end{equation}

We assume that every event reaching the detector is detected so the
$s(\e)$ is given solely by the geometrical constraints
\begin{equation}
s(\e)=\begin{cases}
1  & z_{u}\in [-L/2,L/2] \wedge z_{d}\in [-L/2,L/2]\\ 
0 & \text{otherwise}.
\end{cases}
\end{equation}
This is somewhat more complicated in the image space
\begin{equation}\label{eq:thetalim}
s(y,z,\theta)=\begin{cases}
1  & \tan\theta \in [\max(-\frac{\frac{1}{2}L+z}{R-y},\frac{-\frac{1}{2}L+z}{R+y}),
\min(\frac{\frac{1}{2}L-z}{R-y},\frac{\frac{1}{2}L+z}{R+y})]\\ 
0 & \text{otherwise}
\end{cases}
\end{equation}
We will also need the  sensitivity of the image point $(y,z)$
\begin{equation}\label{eq:s-yz}\begin{split} 
s(y,z) & =\pi^{-1}\int\text{d}\theta \,s(y,z,\theta)=\pi^{-1}(\theta_{max}-\theta_{min})
\end{split}
\end{equation}
with
\begin{equation}
\begin{split}
\theta_{min}&=\arctan\max(-\frac{\frac{1}{2}L+z}{R-y},\frac{-\frac{1}{2}L+z}{R+y}), \\
\theta_{max}&=\arctan\min(\frac{\frac{1}{2}L-z}{R-y},\frac{\frac{1}{2}L+z}{R+y}).
\end{split}
\end{equation}

 As discussed in the previous section the errors are normally distributed
\begin{equation}\label{eq:cond}
p(\tev|\ev)=s(\e)\frac{\det^{\frac{1}{2}} C(\ev)}{(2\pi)^{\frac{3}{2}}}
\exp\left(-\frac{1}{2}(\tev-\ev)^TC^{-1}(\ev)(\tev-\ev)\right)
\end{equation}
where 
\begin{equation}
\de=\ev(z,y,\theta)-\ev(\tz,\ty,\tphi)=
\begin{pmatrix}
z+(\R-y)\tan\theta-\tz-(\R-\ty)\tan\ttheta\\
z-(\R+y)\tan\theta-\tz+(\R+\ty)\tan\ttheta\\
-2y\sqrt{1+\tan^2\theta}+2\ty\sqrt{1+\tan^2\theta}
\end{pmatrix}
\end{equation}

We will now construct an approximation for the formula
\eqref{eq:P}.
We start by calculating 
\begin{equation}\label{eq:p-yz}
p(\tev|y,z)=\pi^{-1}s(y,z)\int\text{d}\theta p(\tev|(y,z,\theta)
\end{equation}

The first approximation we make is to assume that the
correlation matrix $C$ is depending weakly on $e$ and we can
approximate it by its value at $\tev$. 
The integral  \eqref{eq:p-yz} becomes then
\begin{equation}\label{eq:kernel-int}
\begin{split}
  p(\tev|y,z)
  =\pi^{-1}\frac{\det^{\frac{1}{2}} C(\tev) }{(2\pi)^{\frac{3}{2}}}\int\limits_{\theta_{min}}^{\theta_{max}}\!\!\text{d}\theta\,
\exp\left(-\frac{1}{2}(\tev-\ev)^TC^{-1}(\tev)(\tev-\ev)\right).
\end{split}
\end{equation}

We will approximate this integral using the saddle-point
approximation. To this end we first expand the $\de$ in 
\begin{equation}
\dth=\theta-\ttheta
\end{equation}
\begin{equation}\label{eq:de-lin}
\de \approx \vec{o}\dth^2+\vec{a}\dth+\vec{b}
\end{equation}
with
\begin{equation}
\vec{o}=\begin{pmatrix}
-(\dy +\ty -R)\tan\ttheta\cos^{-2}\ttheta\\
-(\dy +\ty +R)\tan\ttheta\cos^{-2}\ttheta\\
-(\dy+\ty)\cos^{-1}\ttheta(1+2\tan^2\ttheta)
\end{pmatrix},
\end{equation}
\begin{equation}
\vec{a}=\begin{pmatrix}
-(\dy +\ty -R)\cos^{-2}\ttheta\\
-(\dy +\ty +R)\cos^{-2}\ttheta\\
-2(\dy+\ty)\cos^{-1}\tan\ttheta
\end{pmatrix}
\end{equation}
and 
\begin{equation}
\vec{b}=
\begin{pmatrix}
\dz-\dy \tan\ttheta\\
\dz-\dy \tan\ttheta\\
-2\dy \cos^{-1}\ttheta
\end{pmatrix}
\end{equation}
where
\begin{equation}
\dy=y-\ty\quad\text{and}\quad \dz=z-\tz.
\end{equation}

After inserting \eqref{eq:de-lin} into the exponent of \eqref{eq:kernel-int} we obtain the expression
\begin{equation}
\begin{split}
\frac{1}{2}\left(\vec{o}\dth^2+\vec{a} \dth +\vec{b}\right)
C^{-1} \left(\vec{o}\dth^2+\vec{a} \dt +\vec{b}\right)
\end{split}
\end{equation}
which we truncate to the quadratic order
\begin{equation}\label{eq:kernel-exp}\begin{split}
\left(\vec{o}C^{-1}\vec{b}+\frac{1}{2}\vec{a}C^{-1}\vec{a}\right)\dth^2+
\vec{a}C^{-1}\vec{b}\dth+\frac{1}{2}\vec{b}C^{-1}\vec{b}.
\end{split}
\end{equation}
After differentiating with respect to $\dth$ we obtain the equation
for the minimum
\begin{equation}
\left(2\vec{o}C^{-1}\vec{b}+\vec{a}C^{-1}\vec{a}\right)\dth+
\vec{a}C^{-1}\vec{b}=0
\end{equation}
with the solution
\begin{equation}\label{eq:dt-min}
\dth_{min}=-\frac{\vec{b}C^{-1}\vec{a}}{\vec{a}C^{-1}\vec{a}+2\vec{o}C^{-1}\vec{b}}.
\end{equation}
Denoting 
\begin{equation}
\tau=\dth-\dth_{min}
\end{equation}
we rewrite the \eqref{eq:kernel-exp} as 
\begin{equation} 
\frac{1}{2}\left(\vec{a}C^{1}\vec{a}+2\vec{o}C^{-1}\vec{b}\right)\tau^2+
\frac{1}{2}\left(\vec{b}C^{-1}\vec{b}-\frac{(\vec{a}C^{-1}\vec{b})^2}{\vec{a}C^{1}\vec{a}+2\vec{o}C^{-1}\vec{b}}\right)
\end{equation}

Finally we obtain
\begin{equation}\label{eq:kernel-approx}
\begin{split}
  p(\tev|y,z)
  &\approx\frac{\det^{\frac{1}{2}} C(\tev) }{(2\pi)^{\frac{3}{2}}}
\exp\left(-\frac{1}{2}\left(\vec{b} C^{-1}\vec{b}
-\frac{\left(\vec{b}C^{-1}\vec{a}\right)^2}{\vec{a}C^{-1}\vec{a}+2\vec{o}C^{-1}\vec{b}}\right)\right)\\
&\phantom{\approx\frac{\det^{\frac{1}{2}}} {(2\pi)}}\pi^{-1}s(y,z)
\int\limits_{\theta_{min}}^{\theta_{max}}\!\!\text{d}\tau\,
\exp\left(
-\frac{1}{2}\tau^2\left(\vec{a}C^{-1}\vec{a}+2\vec{o}C^{-1}\vec{b}\right)
\right)
\end{split}
\end{equation}
and  performing the Gaussian integration we  get
\begin{equation}\label{eq:kernel-final}
\begin{split}
  p(\tev|y,z)
  &\approx\frac{\det^{\frac{1}{2}} C }{2\pi\sqrt{\vec{a}C^{-1}\vec{a}+2\vec{o}C^{-1}\vec{b}}}
\pi^{-1}s(y,z)\\
&\phantom{\approx\frac{\det^{\frac{1}{2}} C(\tev) }{2\pi}}
\exp\left(-\frac{1}{2}\left(\vec{b} C^{-1}\vec{b}
-\frac{\left(\vec{b}C^{-1}\vec{a}\right)^2}{\vec{a}C^{-1}\vec{a}+2\vec{o}C^{-1}\vec{b}}\right)\right).
\end{split}
\end{equation}

We still need to perform the integration over the pixel. We will just
approximated it by the value of \eqref{eq:kernel-final} at its center
\begin{equation}
p(\tev|i)\approx V(i)p(\tev|y_i,z_i)
\end{equation}
and 
\begin{equation}
P(\tev|i)\approx  \frac{p(\tev|y_i,z_i)}{s(y_i,z_i)}
\end{equation}
where $(y_i,z_i)$ denotes the center of pixel $i$.

\section{Validation}

To validate our calculations we compare the formulas \eqref{eq:P} and
\eqref{eq:kernel-final} for few selected events.
The biggest issue here is the estimation of the correlation matrix $C$. 
We will consider the case of diagonal correlation matrix not depending on the positions
\begin{equation}
C^{-1}=
\begin{pmatrix}
\frac{1}{\sigma_z^2} & 0  &  0 \\
0 &     \frac{1}{\sigma_z^2} & 0\\
0 & 0 & \frac{1}{\sigma_{\dl}^2} 
\end{pmatrix}
\end{equation}
From our measurments we estimate
\begin{equation}
\sigma_{z}\approx 10mm,\qquad \sigma_{\dl}\approx 63mm.
\end{equation}

\begin{figure} 
\begin{center}
\includegraphics[width=10cm]{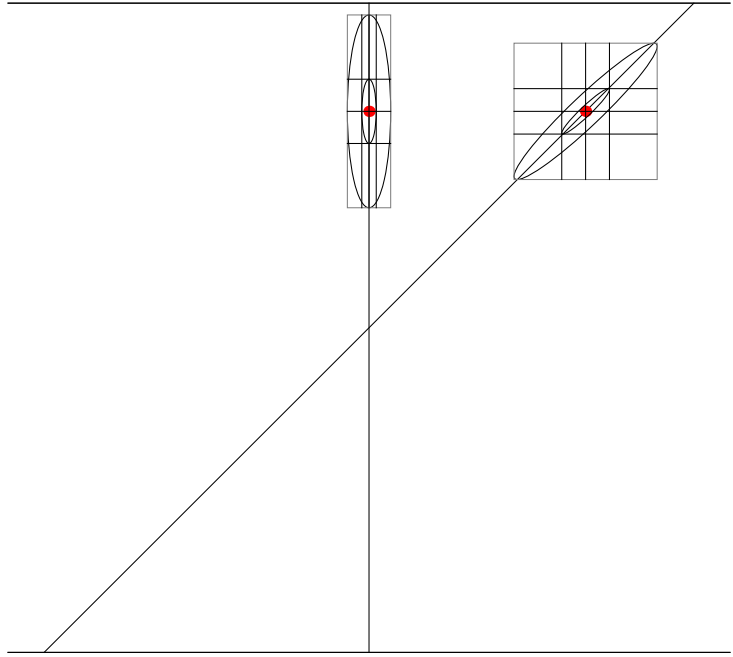}
\end{center} 
\caption{\label{fig:c300} Two of the events used for validation. The
  reconstructed point is at $\ty=300mm$ and reconstructed angles are
  $0^\circ$ and $45^\circ$.}
\end{figure}

We the consider events with $y=300mm$ and angles zero and
$45^\circ$ (see figure~\ref{fig:c300}).  The value
of $z$ does not matter as in this case all the formulas are invariant
with respect to the translation along the $z$ axis.

It is clear that the
formula \eqref{eq:kernel-final} is non-negligible only in a limited
region around the reconstruction point. To estimate this region we
will use only the first term from the exponent.
This a homogeneous polynomial of the second order in $\dy$ and $\dz$
so it defines a ellipse around reconstruction point given by the equation
\begin{equation}
\vec{b}C^{-1}\vec{b}=R^2
\end{equation}
The region of
the interest is defined as the three sigma ellipse  ($R=3$). 
For each event we scan the formulas \eqref{eq:P} and
\eqref{eq:kernel-final} along the horizontal and vertical line
segments based on the bounding box of the one $\sigma$ ellipse (see
the figure~\ref{fig:c300}). For this choice of parameters the two
formulas were practically indistinguishable.

\section{Summary}

We have presented a derivation of the system response kernel for a PET
detector based on time of flight measurements in two parallel
scintillators strips.  The resulting formula for the kernel is still
quite complicated. For each event the expression in the exponent is a
rational function in variables $\dy$ and $\dz$. We could envisage
further simplification, but this can be problematic without detailed
knowledge of the detector geometry/size and the matrix
$C(\e)$. However we believe that our formula provides a very good
starting point for further approximations once the detector geoemtry is fixed.

The biggest simplification we have made is to assume that the
scintillators have no thickness. In reality they can be up to 20mm
thick. A simplest approach would be to incorporate this into the
correlation matrix. However our preliminary calculations show that the
resulting errors are not Gaussian. This is a subject of an ongoing
investigation.\\

\textbf{Acknowledgments}\\
We acknowledge technical and administrative support by M.~Adamczyk, T.~Gucwa-Ry\'{s}, A.~Heczko, 
K.~\L{}ojek, M.~Kajetanowicz, G.~Konopka-Cupia\l{}, J.~Majewski, W.~Migda\l{}, and the financial support 
by the Polish National Center for Development and Research and by the Foundation for Polish Science through MPD programme,
the EU and MSHE Grant No. POIG.02.03.00-00-013/09.

\end{document}